\documentclass[twocolumn]{elsart}
\usepackage{natbib,amssymb,graphicx}
\journal{New Astronomy Reviews}
\begin{document}
\runauthor{Malesani et al.}
\begin{frontmatter}
%\title{The Lorentz factor of gamma-ray bursts measured by REM}
\title{The GRB afterglow onset observed by REM:\\fireball Lorentz factor and afterglow fluence}
\author[DARK]{Daniele Malesani},
\author[Brera]{Emilio Molinari},
\author[DIAS,DCU]{Susanna D. Vergani} \and
\author[Brera]{Stefano Covino}
\author{on behalf of the REM team}
%\thanks[X]{This is the history of the paper, etc etc}

\address[DARK]{Dark Cosmology Centre, Niels Bohr Institute, Juliane Maries Vej 30, DK-2100 K{\o}benhavn \O, Denmark}
\address[Brera]{INAF, Brera Astronomical Observatory, via Bianchi 46, I-23807 Merate (Lc), Italy}
\address[DIAS]{Dunsink Observatory, DIAS, Dunsink lane, Dublin 15, Ireland}
\address[DCU]{School of Physical Sciences and NCPST, Dublin City University, Dublin 9, Ireland}
%\thanks[Someone]{Partially supported by the Roman Senate}

\begin{abstract}
We report observations of the early light curves of GRB\,060418 and
GRB\,060607A, carried out with the pink robotic telescope REM. A clear peak is
detected for both events, which is interpreted as the onset of the afterglow,
that is the time at which the fireball starts decelerating. This detection
allows to directly measure the initial fireball Lorentz factor, which was found
to be $\Gamma_0 \approx 400$ for both events, fully confirming the
ultrarelativistic nature of gamma-ray burst fireballs. Sampling the light curve
before the peak also allows to compute the bolometric fluence of the afterglow,
which is 16\% of the prompt one in the case of GRB\,060418.
\end{abstract}


\end{frontmatter}

\section{Introduction}

It has long been known that the plasma emitting gamma-ray bursts (GRBs) must be
moving relativistically, and that its Lorentz factor $\Gamma$ is much larger
than unity. This follows by the so-called compactness argument
\citep{Ruderman75}. The high photon densities, coupled with the short
variability timescales, imply that GRB sources should be optically thick to
pair production, leading to a huge suppression of the emitted flux and to
thermal spectra, contrary to what is observed. The solution to the compactness
problem requires the source to be in relativistic motion \citep{Piran00}. Lower
limits to the Lorentz factor $\Gamma \gtrsim 100$ are usually derived
\citep{LithwickSari01}.

The discovery of long-lived afterglows has greatly advanced our knowledge of
GRBs. Afterglow radation is powered by the deceleration of the relativistic
fireball. The afterglow behaviour at late times, however, is insensitive to the
initial Lorentz factor, since the fireball decelerates in a self-similar way
\citep{BlandfordMcKee76}. The fireball Lorentz factor can be measured by
observing the afterglow onset \citep{SP99}, which roughly corresponds to the
time at which the fireball starts decelerating significantly. At this time, the
afterglow luminosity reaches a maximum. Unluckily, the early light curves are
very complex, and the observed emission is a mixture of several components,
which easily hide the afterglow peak: residual prompt activity, reverse shocks,
late internal shocks, reverberberation of the main GRB. A clear peak could be
observed in very few cases, most noticeably GRB\,030418 and GRB\,050820A
\citep{Rykoff04,Vestrand06}.

The \textit{Swift} satellite triggered on the long-duration GRB\,060418 and
GRB\,060607A, promptly located them, and for both discovered an X-ray and
optical afterglow (Falcone et al. 2006; Ziaeepour et al. 2006). Their redshifts
are $z = 1.489$ and 3.082, respectively, thus implying an isotropic-equivalent
energy $E_{\rm iso} = 9 \times 10^{52}$ and $\sim 1.1 \times 10^{53}$ erg
\citep{Dupree06,Vreeswijk06}. The X-ray telescope followed their light curve
for a few days, revealing intense flares for both. The REM (Rapid Eye Mount)
robotic telescope \citep{Zerbi01,Chincarini03} promptly reacted to the
triggers, and started observing the GRB fields about one minute after the GRB,
locating in both cases a near-infrared (NIR) counterpart
\citep{Covino06a,Covino06b}. In the case of GRB\,060418, multifilter
observations were secured to study the afterglow spectrum, while for
GRB\,060607A a single, densely sampled light curve was recorded. We refer to
\citet{Molinari07} for a full description of these data.

\begin{figure*}
\centering\includegraphics[width=0.9\textwidth]{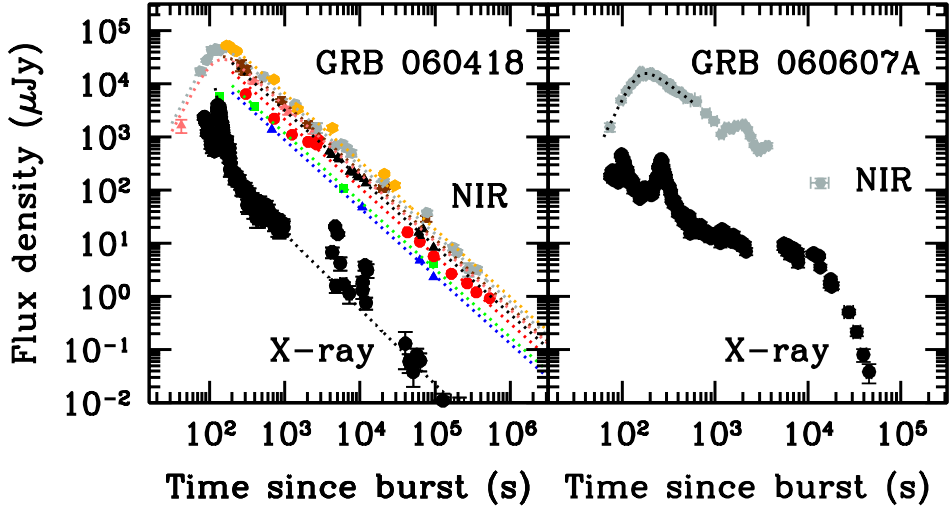}
\caption{X-ray and NIR/optical light curves of GRB 060418 and GRB\,060607A. The
REM data have been complemented by GCN and VLT data.}
\label{fg:lcs}
\end{figure*}

\section{The fireball Lorentz factor}

Figure~\ref{fg:lcs} shows the light curves of GRB\,060418 and GRB\,060607A. In
the NIR, a clear peak is observed $\approx 150$~s after the trigger. Following
the maximum, the curves evolve gradually into a power-law decay. This is
different from what observed in the X rays, where the flares are observed
superimposed to an underlying component with power-law behaviour. For
GRB\,060418, the decay goes on interrupted for more than three decades in time,
directly linking the peak to the forward shock emission. These properties
suggest that the observed maximum corresponds to the afterglow onset. The peak
times were quantitatively determined by fitting a smoothly-broken power law to
the light curve.

The observed peak times $t_{\rm peak}$ (150 and 180 s for GRB\,060418 and
GRB\,060607A, respectively) are longer than the burst durations, and this
corresponds to the so-called thin-shell case. In this scenario, the afterglow
peak time roughly marks the epoch at which the expanding fireball has swept up
enough mass to be significantly decelerated. Using the formulation by
\citet{SP99}, we have
\begin{equation}
  \Gamma_0 =
    320 \left[
      \frac{E_{\gamma,53}(1+z)^3}{\eta_{0.2}n_0t_{\rm peak,2}^3}
    \right]^{1/8},
\end{equation}
where $E_\gamma = 10^{53}E_{\gamma,53}$~erg is the fireball
(isotropic-equivalent) energy, $n = n_0$~cm$^{-3}$ is the particle density of
the surrounding medium (supposed homogeneous), $\eta = 0.2\eta_{0.2}$ is the
radiative efficiency, and $t_{\rm peak,2} = t_{\rm peak}/(100~\mathrm{s})$. We
infer $\Gamma_0 \approx 400$ for both bursts, weakly dependent on the unknown
efficiency and external medium density.

In our computation, we have assumed a homogeneous medium. The light curve
before the peak indeed rises as $\sim t^3$, consistent with the expectations
for a uniform ISM \citep{JinFan07} and in contrast with a wind-shaped ($n
\propto r^{-2}$) environment. After the peak, however, the behaviour of
GRB\,060418 is inconsistent with both a homogeneous and a wind medium. This
might be due, for example, to varying microphysical parameters, or presence of
Compton emission, or radiative losses. Assuming a wind-shaped density profile,
we find a somehow lower value for the Lorentz factor, $\Gamma_0 \approx 150$.

The measured values are in agreement with theoretical predictions and
consistent with existing lower limits \citep{LithwickSari01}. Using $\Gamma_0
\approx 400$, we compute the emission radius $R = 2ct_{\rm peak} [\Gamma(t_{\rm
peak})]^2/(1+z) \approx 10^{17}$~cm. This is much larger than the internal
shocks scale (where the prompt emission is believed to arise), confirming the
different origin of these two components. Albeit $\Gamma_0$ is similar for
GRB\,060418 and GRB\,060607A, a universal value is unlikely. For example, no
peak was observed for GRB\,050401 \citep{Rykoff05}, implying $\Gamma_0 > 900$.

\section{Afterglow energetics}

The detection of the peak allows the measurement of another important quantity,
the afterglow bolometric fluence ${\mathcal F} = \int F_\nu(t,\nu)\,
\mathrm{d}\nu\, \mathrm{d}t$. The integration over the frequency domain
requires also the knowledge of the spectral shape. For GRB\,060418, our
multiwavelength coverage, coupled with the X-ray monitoring, allows to
determine the peak frequency as a function of time, and the spectrum can be
safely extrapolated. The host-galaxy extinction ($A_V = 0.1$~mag) was computed
by imposing for the optical/NIR and X-ray spectral slopes $\beta_{\rm opt} =
\beta_{\rm X} - 0.5$, and assuming an SMC extinction curve.

By computing the integral, we  get ${\mathcal F} = 2.2 \times
10^{-6}$~erg~cm$^{-2}$. To our knowledge, this is the first case for which such
a measurement has been performed. For comparison, the prompt emission
bolometric fluence (easily computed thanks to the broad-band
\textit{Wind}/Konus measurement; \citealt{Golenetskii06}) is ${\mathcal F}_{\rm
GRB} = 1.6\times10^{-5}$~erg~cm$^{-2}$. This implies an afterglow-to-prompt
fluence ratio of 16\%. In principle, external shocks are more efficient in
dissipating the fireball energy than internal collisions (which have a low
Lorentz factor contrast). Our result thus implies that external shocks are not
much efficient in radiating the dissipated energy. This is consistent with the
observed regime of slow cooling inferred by the SED modeling.

DM acknowledges IDA for support. We thank all the collaborators of our work
\citep{Molinari07}.

\end{document}